# Impacts of Quantum Chemistry Calculations on Exoplanetary Science, Planetary Astronomy, and Astrophysics

A white paper submitted in response to Astro2020 call to Astronomy and Astrophysics

Der-you Kao[1,2,*], Marko Gacesa[3,3a], Renata M. Wentzcovitch[4], Shawn Domagal-Goldman[1], Ravi K. Kopparapu[1], Stephen J. Klippenstein[5], Steven B. Charnley[1], Wade G. Henning[6,1], Joe Renaud[7], Paul Romani[1], Yuni Lee[1,10], Conor A. Nixon[1], Koblar A. Jackson[8], Martin A. Cordiner[1,9], Nicholas A. Lombardo[10,1], Scott Wieman[11,1], Vladimir Airapetian[1,12], Veronica Allen[1,2], Daria Pidhorodetska[1,11], Erika Kohler[1,2], Julianne Moses[13], Timothy A. Livengood[6,1], Danielle N. Simkus[1,2], Noah J. Planavsky[14], Chuanfei Dong[15], David A. Yuen[4], Arie van den Berg[17], Alexander A. Pavlov[1], Jonathan J. Fortney[17]

Author Affiliations:
1. NASA Goddard Space Flight Center; 2. Universities Space Research Association; 3. NASA Ames Research Center; 3a. Bay Area Environmental Research Institute; 4. Columbia University, New York, NY; 5. Argonne National Laboratory; 6. University of Maryland, College Park, MD; 7. George Mason University, Fairfax, VA; 8. Central Michigan University, Mt. Pleasant, MI; 9. Catholic University of America, Washington, DC; 10. University of Maryland, Baltimore County, MD; 11. CRESST II, University of Maryland, Baltimore County, MD; 12. American University, Washington, DC; 13. Space Science Institute, Boulder, CO; 14. Yale University, New Haven, CT; 15. Princeton University, Princeton, NJ; 16. Utrecht University, Utrecht, Netherlands; 17. University of California, Santa Cruz, CA

*Corresponding Author: der-you.kao@nasa.gov, 301-614-6710

# Introduction

Several of NASA missions (TESS, JWST, WFIRST, *etc.*) and mission concepts (LUVOIR, HabEx, and OST) emphasize the exploration and characterization of exoplanets, and the study of the interstellar medium. To be able to deduce from these observations the chemical composition of exoplanet atmospheres and the interstellar medium comprehensive knowledge of the relevant photochemistry is required. One-dimensional photochemical models are a key tool in assessing the compositions and climates of a variety of worlds, including rocky and gaseous planets, and planets both in and beyond the Solar System. Such models rely on rate coefficients and reaction product data either measured in the laboratory or estimated using quantum chemical (sometimes called *ab initio* or first principles) transition state theory calculations. These models are then validated against detailed measurements on Earth, or from missions to other planets (*e.g.*, Cassini-Huygens at Titan and Saturn [Vuitton, 2019]). We anticipate that a much broader set of chemical environments exists on exoplanets, necessitating data from a correspondingly broader set of chemical reactions. Similarly, the conditions that exist in astrophysical environments such as the interstellar medium or circumstellar disks, are very different from those traditionally probed in laboratory chemical kinetics studies. These are areas where quantum mechanical theory, applied to important reactions via well-validated chemical kinetics models, can fill a critical knowledge gap.

After years of development of the relevant quantum chemical theories and significant advances in computational power, quantum chemical simulations have currently matured enough to describe real systems with an accuracy that competes with experiments. These approaches, therefore, have become the best possible alternative in many circumstances where performing experiments is too difficult, too expensive, or too dangerous, or simply not possible. In this white paper, several existing quantum chemical studies supporting exoplanetary science, planetary astronomy, and astrophysics are described, and the potential positive impacts of improved models associated with scientific goals of missions are addressed. In the end, a few recommendations from the scientific community to strengthen related research efforts at NASA are provided.

# Current Applications and Impacts

*1. Rate coefficients of species in atmospheres and the interstellar medium:*

<u>I. EXAMPLE</u>: The effect of unvertainties in chemical rates has been studied in the atmospheres of Neptune [Dobrijevic, 1998; Dobrijevic 2010a], Saturn [Dobrijevic, 2003; Dobrijevic, 2011], and Titan [Hebrard, 2013; Dobrijevic, 2014; Loison, 2015]. The reliability of photochemical models depends on the accuracy of rate coefficient inputs. Unfortunately, precise experimental rate constants in low temperature, low pressure environments are difficult to measure [Blitz, 2012].

In recent years, remarkable advances have been made in the ability to predict the kinetics of gas phase chemical reactions with *a priori* theoretical calculations. These advances arise from improved algorithms for electronic structure theory, for dynamics, for kinetics, and for their couplings, as well as from the ever-growing capability of computational resources. The advances enable predicted kinetics of arbitrary gas phase reactions with uncertainties approaching or even exceeding those of experiments. Furthermore, algorithms are now being developed that allow for the automated, high-fidelity prediction of kinetics for thousands of reactions at a time [Cavallotti, 2019; PACChem, 2019]. These algorithms include automated procedures for mapping stationary points on a potential energy surface. Such codes will greatly facilitate the mapping of the chemistry of complex chemical environments, such as those that occur in planetary atmospheres.

Thermochemical properties of moderate-sized molecules can be predicted with accuracies of about 1 kJ/mol [Klippenstein, 2017b]. Similar accuracies are expected for barrier height predictions. Advanced electronic structure methods (e.g., CCSD(T) or double hybrid density functional theory) also provide high accuracy rovibrational properties, allowing for highly accurate partition function predictions. Furthermore, *a priori* methods for predicting the pressure dependence of the reactions are now available. As a result, *ab initio* rate predictions with an accuracy of a factor of two or better are now routine for many reactions under a wide range of conditions. In favorable circumstances, such as barrierless reactions or at high temperatures, accuracies approaching 20% appear achievable [Jasper, 2014; Klippenstein 2017a].

At the very low temperatures relevant to interstellar chemistry, key reactions are generally barrierless due to the small value of the Boltzmann factor for positive barriers. For such barrierless reactions, a two-transition state model has been repeatedly shown to yield quantitative rate predictions even down to low temperatures [see, e.g., Sabbah, 2007]. The two-transition state model [Greenwald, 2005] couples variable reaction coordinate transition state theory [Klippenstein, 1992] (to treat the long-range transition state) with more standard variational transition state methods [Truhlar, 1996] (to treat the shorter-range transition state).

II. IMPACT: The key challenge of exoplanetary science is moving from discovery to characterization. As we learn more about atmospheres, we will build up a clearer picture of each exoplanet's construction, history, and habitability. Photochemical models can interpret detailed chemical molar abundances from observatory spectra of exoplanetary atmospheres (LUVOIR and HabEx are designed for these critical observations). In addition, the evolution of the planets (OST addresses the formations of planets) can be revealed by studying the chemical reactions in the interstellar medium. Accurate rate coefficients are needed to precisely explain surrounding conditions, or pathways to the formation of molecules (more discussion in *4.*). Therefore, accurate chemical rate coefficients are essential for the characterization of exoplanets and the study of our universe. Quantum chemical calculations can also consolidate models more effectively and more economically than laboratory work.

*2. Interior structures of planets:*

I. EXAMPLE: The first results of numerically modelling of mantle convection in large terrestrial exoplanets using a material model based on *ab initio* calculations, using density functional theory, were very recently presented in van den Berg [2019]. The results clearly reveal the need for combining realistic equation of state parameters, together with self-consistent compressible convection simulations. The prevailing pressure (0–3000 GPa) and temperature conditions (1000--5000 K) [Stamenković, 2011] in the deep mantle of such planets are outside the range accessible to actual physical experiments in the laboratory. Therefore, it is necessary to apply theoretical mineral physics models to obtain material properties, including, for example, complex phase transitions.

Materials problems encountered in geophysics and planetary science have motivated fundamental methodological developments in materials theory and simulations [e.g., Wentzcovitch, 1993; Wentzcovitch, 2004; Zhang, 2014]. Conversely, materials simulations have contributed fundamental advances in geophysics and planetary sciences [e.g., Wentzcovitch, 2004; Wentzcovitch, 2006; Wu, 2014a]. For over two decades, *ab initio* materials simulations have been providing predictions of thermodynamic and thermoelastic properties of planet forming phases at thermodynamically relevant conditions that are challenging to experiments. From metallic iron, to planet forming silicates and melts, to water and ices, and fluid hydrogen/helium mixtures (see [Wentzcovitch, 2010] for a review), *ab initio* predictions are filling an essential gap of information needed for planetary modeling. *Ab initio* materials discovery has also been essential for identifying new planet-forming phases, from the identification of the post-perovskite (PPV) phase [Tsuchiya, 2004] of $MgSiO_3$, the highest pressure form of a mantle-forming silicate on Earth, to the discovery of several post-PPV phase transitions [Umemoto, 2006; Umemoto, 2010; Umemoto, 2011; Umemoto, 2017; Wu, 2014b], producing the phases known today as Super Earth's mantle-forming phases.

The discoveries of these materials and their associated thermoelastic properties have enabled more realistic modeling of the internal structure and dynamical state of Earth and super-Earth type planets. Convection simulations in the Earth have evolved to the point that they can use pressure and temperature dependent materials properties [Tosi, 2013; Matyska, 2011]. Such simulations are now also being used to predict the mass dependent internal dynamical state of super-Earth type planets [van den Berg, 2010; Shahnas, 2018; van den Berg, 2019]. Despite these advances, the chemical space of potential planet forming phases explored so far has been limited to the $MgO$-$SiO_2$ (MS) system, which is fundamental, but incomplete for terrestrial type planets that also contain $Al_2O_3$, $FeO$, $CO_2$, and $H_2O$, at least (MAFSCH). Further advances toward a better understanding of these most interesting exoplanets will require investigations of possible phases and states of aggregation in this more extended chemical space at pressures up to 3 TPa [Umemoto, 2017] and temperatures up to ~10,000 K.

Most of the exoplanets are discovered so far are fallen in the hot-Neptune family. Therefore, a lot of scientific studies are centered around them. The interiors of newly discovered exoplanets and the giant ice planets, Uranus and Neptune, is loosely constrained, because the limited observational data can be satisfied with various interior models. The mantles of icy planets comprise large amounts of water, ammonia, and methane ices, but there is not much known about their phases inside the planets. First-principles calculations show that 2:1 ammonia hydrate is stabilized at icy planet mantle conditions [Robinson, 2017]. Above 65 GPa, ammonia hydrate transforms from a hydrogen-bonded molecular solid into a fully ionic phase $O^{2-}(NH_4^+)_2$, where all water molecules are completely deprotonated. Ammonia hemihydrate is stable in a sequence of ionic phases up to 500 GPa, pressures found deep within Neptune-like planets, and thus at higher pressures than any other ammonia–water mixture. This suggests it precipitates out of any ammonia–water mixture at sufficiently high pressures and thus forms an important component of icy planets.

II. IMPACT: Given the physical limitations and high costs of experiments for reaching pressures relevant for even Earth's core and core-mantle boundary, *ab initio* simulations will provide an alternative for studying the structure, evolution, volatile release, and even habitability of terrestrial words in the 2-10 Earth-mass range where yet undiscovered mineral phases may exist. For Neptune-class worlds, the transition between the superionic and plasma phase of water ice [Zeng & Sasselov, 2014] remains poorly constrained, yet is utterly critical to the character of any world in this class, determining whether it tidally dissipates more like an ocean world or solid world, with broad consequences for the stability of moons around such planets.

*3. Photochemical escape:*

I. EXAMPLE: Dissociative recombination of photoionized $O_2^+$ with electrons in the upper atmosphere of Mars produces translationally hot (energies up to ~7 eV in center-of-mass frame) oxygen atoms capable of escaping into space and forming the Martian hot O corona [Lammer, 2008]. This process has been found to be a major escape mechanism presently active on Mars, as well as a key process for explaining the loss of primordial Martian water [Lillis, 2017]. The O escape flux strongly depends on $O+CO_2$ cross sections, as well as those of $O+CO$, $O+H_2$, $O+He$, and others [Fox, 2018]. Moreover, hot O atoms power secondary escape processes by colliding with thermal atmospheric species, and ejecting them to space [Zhang, 2009]. These processes cannot be neglected in the analysis of the isotopic composition of the Martian (and possibly Venusian) atmosphere. The same processes likely occur on exoplanets, and may be particularly relevant for the habitability of sub-Earth mass exoplanets and/or planets in orbit around active stars that provide abundant ionizing radiation.

II. IMPACT: A better understanding of atmospheric escape and the collisions of ionic and neutral molecules, has major implications for the prediction of atmospheric evolution and habitability of planets around active stars. Accurate estimates of the cross sections of neutral-neutral collisions done by *ab initio* calculations will improve atmospheric escape models. Given the potential focus of JWST and ELT on these targets, this is a critical area of study for the next decade's astronomy observations. In addition, spectral line broadening is observed in Venus' atmosphere. Understanding the collisional broadening of line profiles will improve remote sensing and detection capabilities of space-based telescopes. Similarly, collisionally induced absorption (CIA) plays a significant role in understanding the evolution of Mars' atmosphere [Wordsworth, 2016] and is connected to the Faint Young Sun paradox.

*4. Understanding the history of the development of the universe, and life:*

I. EXAMPLE: The isotope fractionation of a molecule dependents on its formation environment and the process by which it was formed. Therefore, isotopic fractionation has been widely used to trace the physical and chemical conditions of environments, and life too.

Isotopic fractionation in the interstellar medium can be driven by multiple mechanisms: isotopic-exchange reactions in the gas phase, isotope selective UV photodissociation, and grain surface reactions. In addition to these local processes, elemental abundance ratios depend on the galactocentric distance and are not constant with time because of stellar nucleosynthesis [Wilson, 1999].

The data needed to investigate isotope fractionations in the interstellar medium are reaction rate constants, which imply possible or impossible chemical pathways. A time-dependent chemical model was built for gas phase ion-molecule and atom-molecule reactions involving D, $^{13}C$, and $^{15}N$ species and chemical network and quantum chemical calculations were applied to determine the possible presence of reaction barriers in the network [Roueff, 2015] due to the low temperature of the environment. The simulation predicts an elemental ratio of $^{14}N/^{15}N$. However, in many other studies, $^{15}N$ enrichment is incorrectly assumed [Wirström, 2018].

Dimethyl ether is one of the most abundant interstellar organic molecules, but its formation route remained controversial until a very recent quantum chemical kinetic study determined the rate coefficient of interstellar dimethyl ether gas-phase formation and thereby identified the environmental conditions necessary for its formation [Skouteris, 2019].

Isotope fractionations of metal centered coordination compounds are used to determine the oxidation processes in Earth's history [Domagal-Goldman, 2009; Saad, 2017]. The redox environment of Earth is closely associated with the emergence of life, which in turn, via biochemical reactions, also causes isotopic fractionation. Quantum chemical simulations have been employed in these studies to factor out surrounding interference, that are otherwise difficult to control in the laboratory. Isotope fractionation can be calculated based on vibrational frequencies, which are easily done in quantum chemical calculations. In addition, chemical species in space are often identified using vibrational spectra and that quantum chemical calculations can decipher the types of molecules present in space. A polycyclic aromatic hydrocarbon (PAH) infrared spectral database, built by Ames Research Center, has been used to disentangle the observed spectrum into contributions from various molecules in the photodissociation region [Rosenberg, 2011].

II. IMPACT: Searching for life will be one of NASA's research concentrations in the next decade, involving a number of missions including TESS, JWST, LUVOIR, HabEx, and OST. An improved understanding of the environment and evolution of ancient and present Earth aided by quantum chemical simulations can build a solid foundation of knowledge on which to base the exploration of exotic worlds.

*5. Surface treatment and contamination control of space telescopes and instruments:*

Quantum chemistry calculations, have been successfully employed to study surface chemistry in applications such as anticorrosion [Somorjai, 2011]. Surface treatments of the mirrors of space telescopes can be very critical because the requirements of wide bandwidth and long operation time. Owing to the oxygen rich environment, a protective layer must be applied after the deposition of the base coating and before air contact in order to prevent oxidation. It is almost certain that the reflective coating of LUVOIR will be aluminum due to the requirement of high reflection over a broad wavelength range. A few protective materials have been studied to fulfill the criteria of easy removal after the telescope enters space [Allred, 2017].

A second concern related to telescopes and instrument design is understanding how cold telescopes/instruments can lead to adsorption of contaminants ($H_2O$, organics, *etc.*) on mirror surfaces. This can then degrade the observational throughput of those telescopes and instruments. For the next generation space telescopes (LUVOIR, HabEx, and OST), the missions are aimed to explore colder targets at near infrared wavelengths and that requires the mirrors and scientific instruments to be cooled to a few tens of Kelvin in order to reach detectable signal-to-noise ratios. However, organic molecules (mostly due to outgassing in the extremely high vacuum environment) can condense on mirror surfaces. Therefore, most of the space telescopes heat up their primary mirrors to evaporate the contaminant molecules, but this heating compromises sensitivity. Simulations of the condensation process may provide an economical means for helping missions study the trade-offs between different choices of mirror surfaces or coatings and the operational temperatures required for an observatory or instrument bay. Quantum chemical calculations can be used to study the bond energy between interesting molecules and the surfaces of mirrors. The missions of LUVOIR and HabEx are intended to study the spectral signatures of exoplanets, requiring high signal-to-noise ratio for this purpose. Quantum chemical calculations are the cheapest approaches for determining optimized cleaning protocols, heating systems, and surface treatments.

**Recommendations**

Missions like the JWST continue the NASA tradition of exploration, probing our universe ever more deeply and laying bare more of its secrets. Based on fundamental physical principles, quantum chemical calculations and simulations represent important tools to aid the discovery and analysis of new phenomena by these missions. Calculations can fill critical knowledge gaps required to understand conditions on distant exoplanets. They can also support basic science and engineering development in areas such as more efficient batteries and quantum computing that will help power future missions. Interdisciplinary collaborations involving quantum chemists should be strongly encouraged. Therefore, we recommend that the committee (1) increase funding opportunities for mission-related quantum chemical calculations, and (2) support joint workshops for physical chemists, astronomers, and geoscientists. We also encourage the quantum chemistry community to document and publish data systematically, to make it available for broader use.